\def \cc{$\rm{cm}^{-3}$}

\def \g{~\rm{g}}

\def \erg{~\rm{erg}}

\def \yr{~\rm{yr}}

\documentclass[12pt,preprint]{aastex}

\usepackage{natbib}
\usepackage{epsfig}
\begin{document}

\title{VIOLENT STELLAR MERGER MODEL FOR TRANSIENT EVENTS}

\author{Noam Soker\altaffilmark{1},
and
Romuald Tylenda\altaffilmark{2}}

\altaffiltext{1}{Department of Physics, Technion-Israel Institute of
Technology,  32000 Haifa, Israel;
soker@physics.technion.ac.il.}

\altaffiltext{2}{Department for Astrophysics, N.Copernicus Astronomical Center,
Rabia\'nska 8, 87-100 Toru\'n, Poland;
tylenda@ncac.torun.pl}

\begin{abstract}

We derive the constraints on the mass ratio for a binary system to merge
in a violent process.
We find that the secondary to primary stellar mass ratio should be
$0.003 \la (M_2/M_1) \la 0.15$.
A more massive secondary star will keep the primary stellar envelope in
synchronized rotation with the orbital motion until merger occurs.
This implies a very small relative velocity between the secondary star and
the primary stellar envelope at the moment of merger, and therefore very weak
shock waves, and low flash luminosity.
A too low mass secondary will release small amount of energy, and will expel small
amount of mass, which is unable to form an inflated envelope. It can however
produce a quite luminous but short flash when colliding with a low
mass main sequence star.

Violent and luminous mergers, which we term {\it mergebursts,}
can be observed as V838 Monocerotis
type events, where a star undergoes a fast brightening lasting days to months,
with a peak luminosity of up to $\sim 10^6 L_\odot$ followed by a slow
decline at very low effective temperatures.

\end{abstract}


\section{INTRODUCTION}

Stellar mergers have been recognized for a long time as events which can be
important for evolution of binary systems. In discussions of globular clusters
stellar mergers are usually considered as the most probable source of blue
stragglers, e.g. De Marco et al. (2005, and earlier
references therein; Sills et al. 2005).
In some cases three or more stars might be involved
(Knigge et al. 2006).
However, the main interest in these cases has been directed toward understanding
the nature of the final product of a merger in terms of its mass, chemical structure
and farther evolution. Little attention, if any, has usually been paid to direct
observational appearances of these events.
This was obviously due to the common belief that the
stellar mergers are very rare events and that, consequently, there is very little
chance to observe them.
However, the discovery of the eruption of V838~Mon in 2002 (Brown 2002)
and subsequent studies of its observed evolution (Munari et~al. 2002;
Kimeswenger et~al. 2002; Crause et~al. 2003; Kipper et~al. 2004; Tylenda 2005),
as well as, of other similar objects, i.e. V4332 Sgr (Martini et~al. 1999;
Tylenda et~al., 2005) and M31~RV (Mould et~al. 1990) have led to suggestions
that these observed events were likely to be due to stellar mergers
(Soker \& Tylenda 2003, Tylenda \& Soker 2006, hereafter TS06).
Likewise an analysis done by Bally \& Zinnecker (2005) shows that stellar
mergers in cores of young clusters, which might be one of channels for
producing very massive stars, can be source of luminous and spectacular
observational events.

Different reasons can lead to stellar mergers. In dense stellar systems, as globular
clusters or cores of young clusters, direct collisions of two stars or interactions
of binaries with other cluster members can quite easily
happen, often leading to a merger (for recent papers and more references see,
e.g., Lombardi et al. 2003; Fregeau et al. 2004; Mapelli et al. 2004; Dale \& Davies 2006;
for a review of these merger possibilities see Bailyn 1995).
In multiple star systems dynamical
interactions between the components or encounters between the system and other stars
can destabilize stellar trajectories so that two components collide and merge.
A binary stellar system can lose angular momentum during its evolution, e.g. due to mass loss,
so the separation of the components decreases, which may finally lead to a merger.
In the latter case, the merger is probably often relatively gentle and does not
lead to spectacular events. This happens when the system reaches and keeps
synchronization until the very merger. The relative velocity between the matter
elements from different components is then very low, there is no violent shock heating
and the orbital energy is released on a very long time scale. However when the binary
component mass ratio is low the secondary is unable to maintain the primary
in synchronization so the so-called Darwin instability sets in and the merger
takes place with a large difference between the orbital velocity of the secondary
and the rotational velocity of the primary. In this case the merger is expected to
be violent, at least in the initial phase when the large velocity differences
are dissipated in shocks. In this paper we analyze this possibility in more
detail and discuss observational appearances of violent mergers triggered by
the Darwin instability in binaries.

We can schematically distinguish between three basic types of merger
events.
Of course, there is a continuous variation between the three types,
but it is instructive to make these three ideal classes.

(1) The secondary is disrupted during the collision, and contributes most of
the mass in the inflated envelope. This is likely to happen in a close to grazing
collision with a not too-compact companion in a very eccentric
orbit. We suggest this type of merger
as explanation for V838 Mon (TS06).

(2) Merger in a binary system which reaches merger in an unsynchronized
rotation (the spin of the primary star is not synchronized with the orbital angular
frequency), and where the secondary survives the initial merger stages.
The secondary then spirals-in inside the primary envelope.
The inflated envelope comes mainly from the primary mass.
The conditions for the occurrence the type of merger are studied
in this paper.

The above two merger types are expected to be violent. The third
type is a non-violent merger:
\newline
(3) Merger between two stars having
a synchronized rotation. The secondary is massive enough to
maintain synchronized orbital motion until merger occurs. The
secondary survives, and the 2 stars form a massive star in a
relatively gentle process.
{{{Although this process is termed non-violent by us, it might still evolve
on a dynamical time scale at some phases, and it has many interesting properties.
The spiraling-in process of the secondary deep inside the envelope will release
a huge amount of orbital energy, which might result in highly distorted
mass loss event (Morris \& Podsiadlowski 2006).
On a later time, the process can alter the evolution of the star on the
HR diagram (e.g., Podsiadlowski et al. 1990).
However, we don't expect a bright flash in these cases. }}}

The gravitational and kinetic energy of the merging binary system
can result in the following observational events:
\begin{enumerate}
\item Flash of light. This flash is formed by emission from a strongly shocked gas,
in the primary and/or secondary envelope, and will be observed as a flash
lasting as long as the secondary is violently slowed-down in the outer regions of
the primary star. This can be from few times the dynamical time of the system
up to a very long time, depending on the condition of the merging system.
For the flash to be bright, the duration should be short, which implies a large
relative velocity between the secondary and rotating primary envelope.
\item Gravitational and kinetic energy of matter expelled to large distances,
and even leaving the system.
The matter that does not leave the system, falls back on a dynamical time scale
at its maximum distance, and when it becomes optically thick it contracts
on its Kelvin-Helmholtz time scale.
The Kelvin-Helmholtz time scale of the inflated envelope is much shorter than
the Kelvin-Helmholtz time scale of the primary star, due to the very high
luminosity and very low mass of the inflated envelope.
The large energy in the inflated envelope and its relatively short contraction time
implies that the energy deposited in the inflated envelope results in a bright
phase of the merging system, lasting much longer than the initial flash.
\item Gravitational energy of the expanding inner layers of the primary
and/or secondary (even destroying the secondary).
This will be largely so when the secondary penetrate the deep layers
of the primary star. When the inner layers of the primary finally relax to
equilibrium it will be on a very long Kelvin-Helmholtz time scale.
\end{enumerate}
To form a bright transient event only the first two energy
channels are relevant.
These two channels require violent interaction between the secondary and primary
star.
In the present paper we study the conditions for a violent merger triggered
by the Darwin instability in a binary system.

\section{VIOLENT MERGER}
We consider a binary system composed of a primary star having a mass
$M_1$ and a radius $R_1$, and a much lower mass secondary $M_2 \ll M_1$.
As the system evolves the two stars can merge.
The merging can be caused by the expansion of the primary as it evolves
along the main sequence and beyond, or by losing orbital angular momentum
via tidal interactions, or both.
In evolved binary systems the wind carries most of the angular momentum.
Young binary systems can interact with remnants from the progenitor disk
and cloud, or with a tertiary object in the system.
We assume that the binary system reaches synchronization before merging.
Then, as the ratio of orbital separation to primary radius decreases
due to one of the processes listed above, the system becomes unstable to
the Darwin instability and merges in a time scale set by tidal interaction.
This time scale is shorter than the evolution time of the system.
We now derive the conditions for this instability, and consider the
consequences.

We make the following assumptions:
\begin{enumerate}
\item A circular orbit.
An eccentric orbit will yield a somewhat more violent merger, as the periastron
orbital speed is higher than the Keplerian velocity at the same radius.
\item The system reaches synchronization (co-rotation). Namely, the orbital angular
velocity is equal to the primary's spin angular velocity.
\item After the Darwin instability starts, the spiraling in process is relatively
fast, such that no angular momentum is lost from the system.
In reality, some angular momentum will be lost, slowing down the
primary's angular rotation (spin).
\item The primary rotation profile is that of a solid body.
In reality, the tidal interaction will spin more the outer regions,
implying higher primary angular velocity.
\item The primary star maintains its spherical structure.
In reality the primary equatorial radius will increase,
slowing down its rotation.
\end{enumerate}

The Darwin (Darwin 1887) instability occurs when the secondary
star cannot maintain anymore the primary in synchronization.
Namely, as the secondary spirals-in, e.g., because of tidal
interaction, the orbital angular velocity is higher than the
primary's rotation angular velocity. The condition for the
instability is (e.g., Eggleton \& Kiseleva-Eggleton 2001) $I_{\rm
orb} < 3 I_1$, where $I_{\rm orb}$ is the moment of inertia of the
binary system, and $I_1 = \eta M_1 R_1^2$ is the moment of inertia of
the primary; $\eta \simeq 0.05$ for main sequence stars of $M_1
\simeq 3 M_\odot$ and only weakly depends on the stellar mass
(Meynet et al. 2006).
Substituting $M_2 \ll M_1$ in the
expression for the orbital moment of inertia,  gives the orbital
separation $a_c$ below which the Darwin instability exists
\begin{equation}
\frac{a_c}{R_1} = \left( \frac{3 \eta}{q} \right)^{1/2}= 2.2  
\left( \frac{\eta}{0.05} \right)^{1/2}
\left( \frac{q}{0.03} \right)^{-1/2},
\label{ac}
\end{equation}
where $q = M_2/M_1$.
The value of $a_c/R_1$ as a function of $q$ and for $\eta=0.05$
is drawn on the upper panel of Figure {\ref{fprop}.
\begin{figure}
{\includegraphics[scale=0.90]{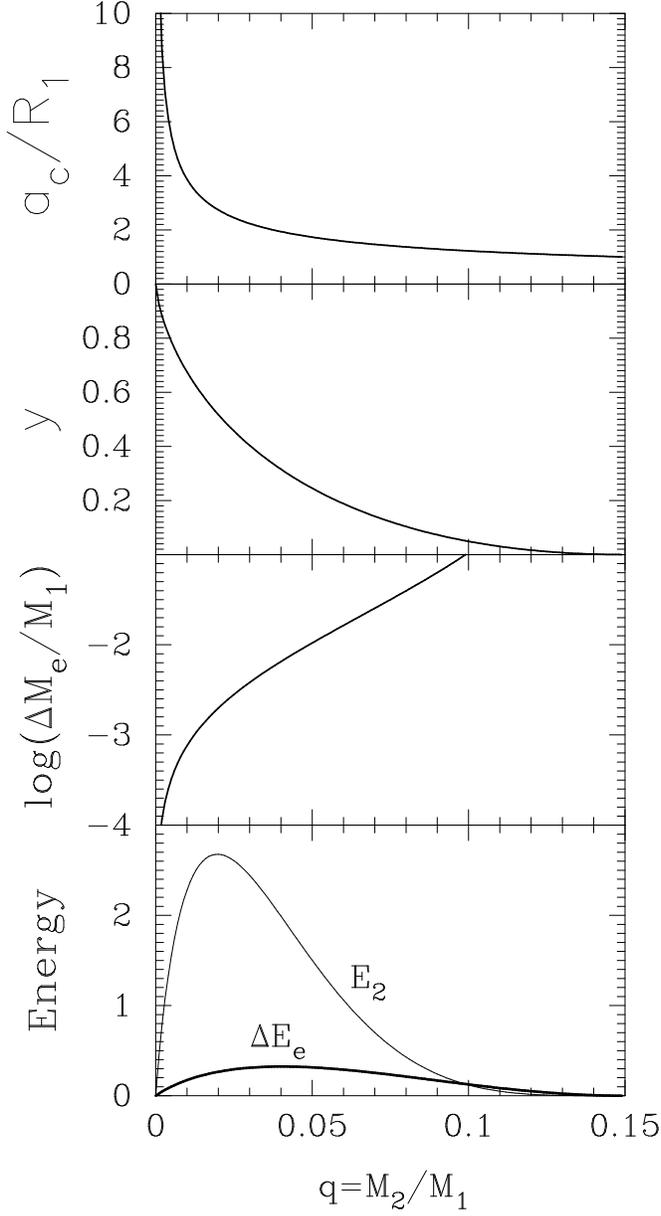}}
\caption{Some quantities as a function of the mass ratio, $q$,
for $\eta=0.05$.
Upper panel: The orbital separation $a_c$, in units
of the primary radius $R_1$, at which the Darwin instability
starts (Eq. \ref{ac}).
Second panel: The relative velocity between the secondary and
envelope as merger starts, in units of $v_{\rm orb}$
(Eq. \ref{vorb1}).
Third panel: The mass that is disturbed at the very outer part
of the envelope, as given by equation (\ref{dme}) with $f=0.9$
and in units of $M_1$.
Lower panel: The secondary kinetic energy (Eq. \ref{en2}) depicted with
a thin line, while the energy deposited in the very outer
part of the envelope, $\Delta R=0.1R_1$ (Eq. \ref{dee}), is drawn with a thick line.
Energies are in units of $10^{-3} GM_1^2/R_1$. }
\label{fprop}
\end{figure}

Under the assumption of a circular orbit the secondary star
enters the primary's envelope with an orbital velocity $v_{\rm orb}$
and angular velocity $\omega_{\rm orb}$ that are given by
\begin{equation}
v_{\rm orb} = \omega_{\rm orb} R_1=\left( \frac{G M_1}{R_1} \right)^{1/2}.
\label{vorb}
\end{equation}
The primary's angular velocity $\omega_1$ and surface equatorial velocity $v_1$
during merger are derived from angular momentum conservation.
Namely, the orbital angular momentum released by the secondary as it spirals-in
from $a=a_c$ to $a=R_1$ equals the change in the primary's angular momentum.
Under the assumptions listed above, we derive
\begin{equation}
\frac{\omega_1({\rm merge})}{\omega_{\rm orb}} =
4 \left( \frac {R_1}{a_c} \right)^{3/2}
-3 \left( \frac {R_1}{a_c} \right)^{2}.
\label{omega1}
\end{equation}
The orbital velocity of the secondary relative to the rotating primary
envelope as merger starts is
\begin{equation}
y \equiv \frac{(v_{\rm orb}-v_1)}{v_{\rm orb}} =
\left[ 1- 4 \left( \frac {R_1}{a_c} \right)^{3/2}
+3 \left( \frac {R_1}{a_c} \right)^{2} \right] .
\label{vorb1}
\end{equation}
The value of $y$ as function of $q$
is drawn in the second panel of Figure {\ref{fprop}.

We can build several other interesting quantities.
First we can derive the secondary kinetic energy relative to the rotating primary envelope
\begin{equation}
E_2 = \frac{1}{2} M_2 (v_{\rm orb}-v_1)^2 =
      \frac{1}{2}\, \frac{G M_1^2}{R_1}\, q\, y^2,
\label{en2}
\end{equation}
This is plotted by the thin line in the lower panel of Figure {\ref{fprop}.
The energy is given in units of
$10^{-3} G M_1^2/R_1 = 3.8 \times 10^{45}(M_1/M_\odot)^2(R_1/R_\odot)^{-1}\erg$.

We are interested in the violent interaction of the secondary with the
primary envelope during merger.
A rather compact secondary (low-mass main sequence star, brown dwarf)
can penetrate quite deeply into the envelope. In this case most of the
event would result in observational effects on a rather long thermal scale.
Also, when energy is deposited well inside the envelope, most
of the energy is channelled to uplift outer envelope layers, causing
only a small observational signature.
However, even in a case like this the initial interaction in the outer
envelope is expected to give violent dynamical effects.
Therefore, we are looking at the interaction, when the
secondary spirals a distance $\Delta R$ in from $a=R_1$ to $a=fR_1$,
where $1-f=(\Delta R/R_1)  \ll 1$.
Let the secondary transfer the angular momentum to an envelope mass
$\Delta M_e$ as it spirals in.
The amount of angular momentum transferred to this mass is
$\Delta J_e \simeq \Delta M_e (v_{\rm orb}-v_1) R_1$, while the
angular momentum lost by the secondary is
$\Delta J_2 \simeq M_2 (G M_1 R_1)^{1/2} (1-f^{1/2})$.
Conservation of angular momentum then gives
\begin{equation}
\Delta M_e \simeq (1 - f^{1/2})\, y^{-1} M_2
  \simeq \frac{\Delta R}{2 R_1} \, y^{-1} M_2 .
\label{dme}
\end{equation}
This mass is actually the mass that was strongly shocked
and is free to expand with almost no disturbances.
A large fraction of this mass is expected to be expelled from the system
and/or to form an extended envelope due to its large entropy.
The value of $\Delta M_e$ as function of $q$, in units of $M_1$ and
for $f=0.9$, is given in the third panel
of Figure {\ref{fprop}.
Its derivation assumes local interaction of the secondary star with the
primary envelope, and therefore it is applicable only to low mass secondary
stars, $q < 0.1$; for more massive secondary stars the value of $\Delta M_e$
as given above requires non-local interaction.

The energy carried by the mass $\Delta M_e$ is
\begin{equation}
\Delta E_e \simeq \frac{1}{2} \Delta M_e (v_{\rm orb}-v_1)^2 =
  \frac{{1-f^{1/2}}}{2}\, \frac{G M_1^2}{R_1}\, q\, y
  \simeq \frac{\Delta R}{4 R_1} \frac{G M_1^2}{R_1}\, q\, y.
\label{dee}
\end{equation}
This is plotted for $\Delta R=0.1 R_1$ by the thick line in the lower panel of
Figure {\ref{fprop}; the energy is given in units of $10^{-3} G M_1^2/R_1$.
This energy is actually the energy residing in gas with high entropy
that is expected to be expelled and/or form the extended post-merger
envelope.

 From the results plotted in Figure {\ref{fprop} alone we can deduce that
for an efficient eruptive (violent) merger the mass ratio should be
in the range $0.003 \la q \la 0.1$.
Massive companions can liberate a huge amount of energy as they
merge.
However, because the synchronization is kept till the merger, the
process is less violent. The system will not be observed as a violent
luminous transient event. Low mass companions ($q \la 0.003$) release
little energy resulting in a rather weak event,
although in some cases (see next section) they can give rise
to short but luminous bursts.

We can estimate the rate at which the energy is released during
the violent phase of the merger and the time scale of this event.
To do so we have to estimate
an effective interaction radius, $R_{\rm eff}$, of the secondary spiraling inside
the primary's envelope. One can do this in an analogous manner as defining the
Bondi-Hoyle accretion radius. In our case the primary's matter moving relative to
the secondary is in the gravitational potential of the primary, so we define
$R_{\rm acc}$ as satisfying
\begin{equation}
  \frac{G M_2}{R_{\rm acc}} \simeq \frac{1}{2} (v_{\rm orb} - v_1)^2 +
                        \frac{G M_1}{R_1} =
     \frac{G M_1}{R_1} \left( \frac{1}{2}\, y^2 + 1 \right)
\label{r_acc}
\end{equation}
The term $(1/2)\,y^2$  is never grater than 0.5,
so we can neglect it in a first approximation, and obtain
\begin{equation}
  R_{\rm acc} \simeq q\, R_1.
\label{r_acc_2}
\end{equation}
In a number of cases, e.g. a brown dwarf merging with a solar type
star,  Eq.~(\ref{r_acc_2}) predicts $R_{\rm acc}$ lower than the radius
of the secondary, $R_2$. In a case like this it seems to be more reasonable to adopt
$R_{\rm eff} = R_2$ instead of $R_{\rm eff} = R_{\rm acc}$.
For a general purpose we therefore adopt
\begin{equation}
  R_{\rm eff} = \xi\, R_1,
\label{r_eff}
\end{equation}
where
\begin{equation}
  \xi = {\rm max} (q,\ R_2/R_1).
\label{xi}
\end{equation}

The rate of the energy dissipation, $L_{\rm diss}$, can be estimated from
\begin{equation}
  L_{\rm diss} \simeq \frac{1}{2}\, \pi\, R_{\rm eff}^2\,
        \rho_1\ (v_{\rm orb} - v_1)^3
     = \frac{1}{2} \pi\, \xi^2\, y^3\, (G M_1)^{3/2}\, R_1^{1/2}\, \rho_1,
\label{lum_d}
\end{equation}
where $\rho_1$ is the density in the primary's envelope.
Note that if a significant part of the dissipated energy goes for producing an
inflated envelope (as discussed below) and/or for mass loss the radiation luminosity
will be significantly lower than the above estimate of $L_{\rm diss}$.

The time scale of the violent merger can be estimated from
\begin{equation}
  \tau \simeq \frac{\Delta E_e}{L_{\rm diss}}
    \simeq \frac {(1 - f^{1/2})\, M_1^{1/2}\, q}
                 {\pi\, G^{1/2}\, R_1^{3/2}\, y^2\, \xi^2\, \rho_1} .
\label{tau1}
\end{equation}
Note that Eq. (\ref{tau1}) estimates the time scale of the initial violent phase when
the shocked matter can easily escape from below the primary's surface and become observable.
The total merger can last significantly longer depending on how deep the secondary can
penetrate before being disrupted and how much time the energy released during the
final phases would require to diffuse to the photosphere.

A significant part of the energy liberated during the merger event can go
to produce an inflated envelope.
To estimate the radius of the envelope, $R_{ie}$, we follow TS06.
The envelope is taken to be a $n=3$ polytropic gas (Tylenda 2005)
siting on top of the merger product in a quasi hydrostatic equilibrium,
and with a total mass $M_{ie}=\beta \Delta M_e$.
The gravitational energy of the inflated envelope is (TS06)
\begin{equation}
E_{ie} \simeq - \frac{G M_1 \beta \Delta M_e}{2 R_1} \frac{1}{\ln(R_{ie}/R_1)}.
\label{eie}
\end{equation}
The energy of this mass prior to merger was
\begin{equation}
E_{ie0} \simeq \frac{G M_1 \beta \Delta M_e}{R_1}
\left[-1+\frac{(1 - y)^2}{2} \right].
\label{eie0}
\end{equation}
We assume that most of the orbital energy liberated
as the secondary spirals-in from $a=R_1$ to $a=fR_1$ goes to
inflate the envelope.
Therefore, the difference in the energy of the inflated mass
before and after merger is about equal to the energy liberated
by the spiraling-in secondary star
\begin{equation}
E_{ie}-E_{ie0} \simeq \frac{G M_1 M_2}{2 R_1}\left(\frac{1}{f}-1\right).
\label{encon}
\end{equation}
Equations (\ref{eie})-(\ref{encon}) can be solved to read
\begin{equation}
  \ln \frac{R_{ie}}{R_1} \simeq \left[2-
     (1 - y)^2 -
     \frac{M_2}{\beta \Delta M_e} \left(\frac{1}{f}-1\right) \right]^{-1}.
\label{lnr1}
\end{equation}
We use equation (\ref{dme}) to substitute for $M_2/\Delta M_e$,
take $f=1-(\Delta R/R_1)$ with $\Delta R \ll R_1$, and neglect high order
in $\Delta R/R_1$. We derive
\begin{equation}
\ln \frac{R_{ie}}{R_1} \simeq
  \left[ 1 - 2 y \left( \frac{1}{\beta} - 1 \right)
     - y^2 \right]^{-1}.
\label{lnr2}
\end{equation}

%
\begin{figure}
{\includegraphics[scale=0.90]{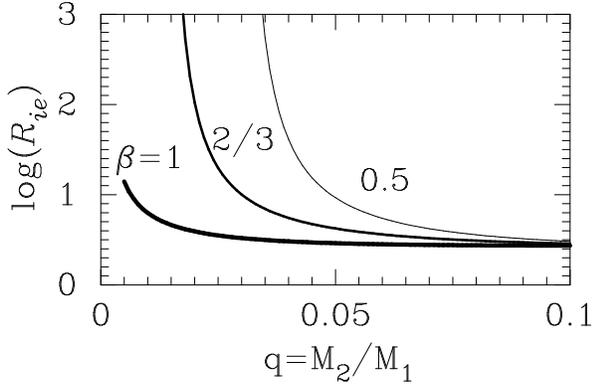}}
\caption{Logarithm of the radius of the inflated envelope in
units of $R_1$, as given by equation (\ref{lnr1}) or (\ref{lnr2}).
$\beta$ is the mass in the inflated envelope in units of $\Delta M_e$
(see Eq. \ref{dme}).}
\label{fien}
\end{figure}

Equation (\ref{lnr2}) is a crude estimate of the radius of
the inflated envelope. From Figure \ref{fien} we can see that if a
fraction $\beta \la 0.7$ (depending on $q$) of the shocked mass $\Delta M_e$
is ejected to the inflated envelope, then a radius of $R_{ie} \sim few \times 100 R_1$
can be achieved (as long as the fraction is not too low).

\section{THE PRIMARY STAR}

A violent merger induced by the Darwin instability can occur if the mass
ratio $q \la 0.1$ (see e.g. Fig.~\ref{fprop}). Details of the event, e.g.
luminosity and time scale, depends on the masses and structures of the
merging components. To show this let us consider three cases, in which the primary,
i.e. more massive of the merging companions, is a $1\,M_\odot$ main sequence
star (case 1), an $8\,M_\odot$ main sequence star (case 2), and a $3\,M_\odot$
red giant (case 3).

In case 1 $q \la 0.1$ means that the secondary is a very low mass star, a
brown dwarf or a massive planet. In all these cases the radius of the
secondary is $R_2 \simeq 0.1\,R_\odot$. This is greater than $R_{\rm acc}$
defined by Eq.~(\ref{r_acc_2}) so we assume $\xi \simeq 0.1$. We can also take
$\rho_1 = 3 \times 10^{-3}$\g\,\cc as a typical density at $0.95-0.90\, R_1$.

The main sequence lifetime of an $8\,M_\odot$ star is $\sim 3 \times
10^7$\,years. This is shorter than the pre-main-sequence lifetime of a
secondary $< 1\,M_\odot$. Using Eq.~(A.4) in TS06, we
obtained, for an age of $1 \times 10^7$\,years, $R_2 \simeq 0.7
(q/0.03)^{2/3}\,R_\odot$. For $q < 0.1$ this is larger than $R_{\rm acc}$,
so in case 2, assuming $R_1 = 5\,R_\odot$, we can take $\xi \simeq 0.14
(q/0.03)^{2/3}$. As the density in the primary envelope we assume
$\rho_1 = 3 \times 10^{-5}$\g\,\cc.

As parameters of the red giant primary in case 3 we take $M_1 = 3\,M_\odot$,
$R_1 = 30\,R_\odot$, and $\rho_1 = 5 \times 10^{-6}$\g\,\cc.
Down to $q \simeq 0.003$ the radius of the secondary is smaller than
$R_{\rm acc}$ (Eq.~\ref{r_acc_2}), so we take $\xi \simeq q$.

\begin{figure}
{\includegraphics[scale=0.40]{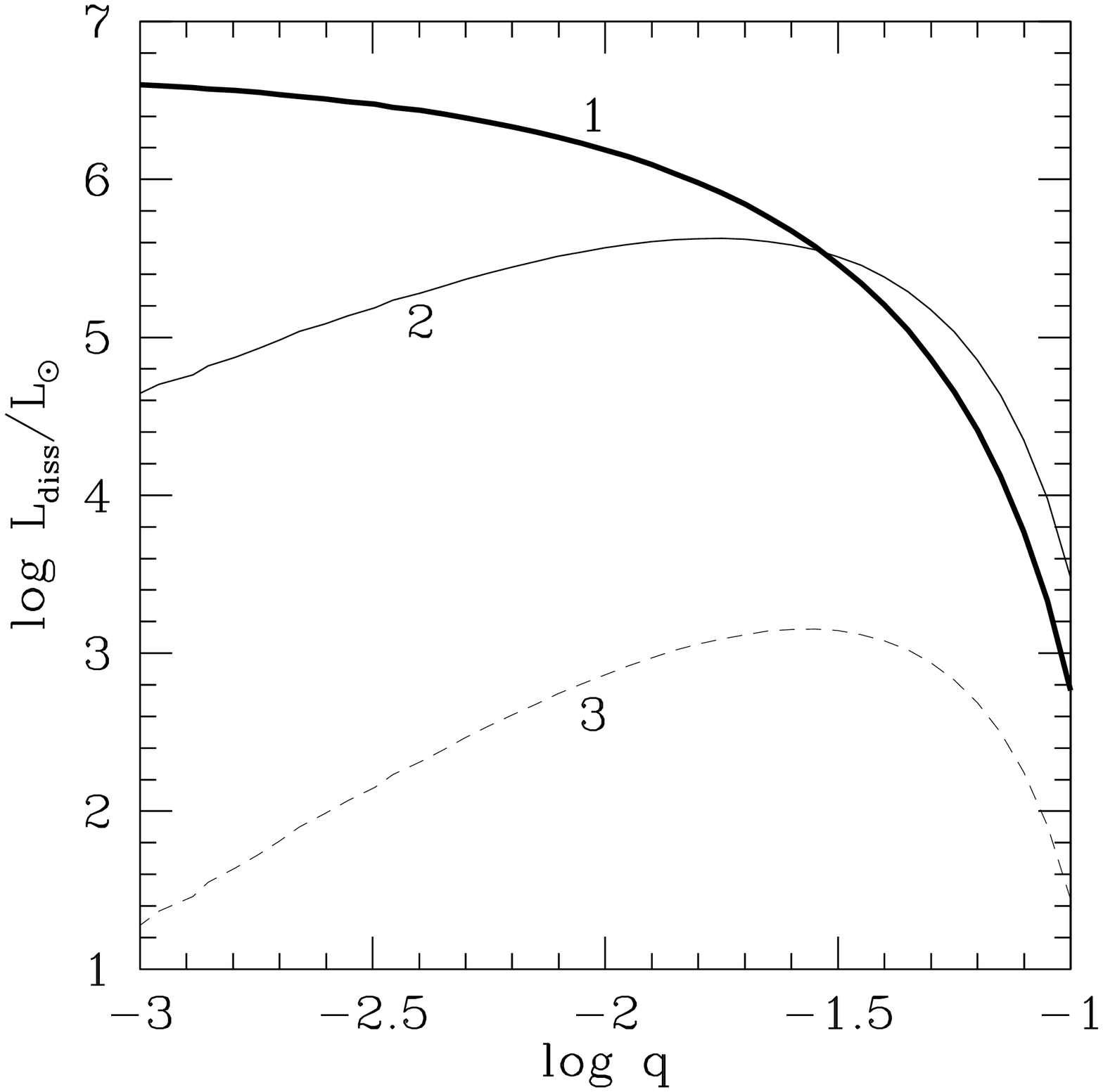}}
{\includegraphics[scale=0.40]{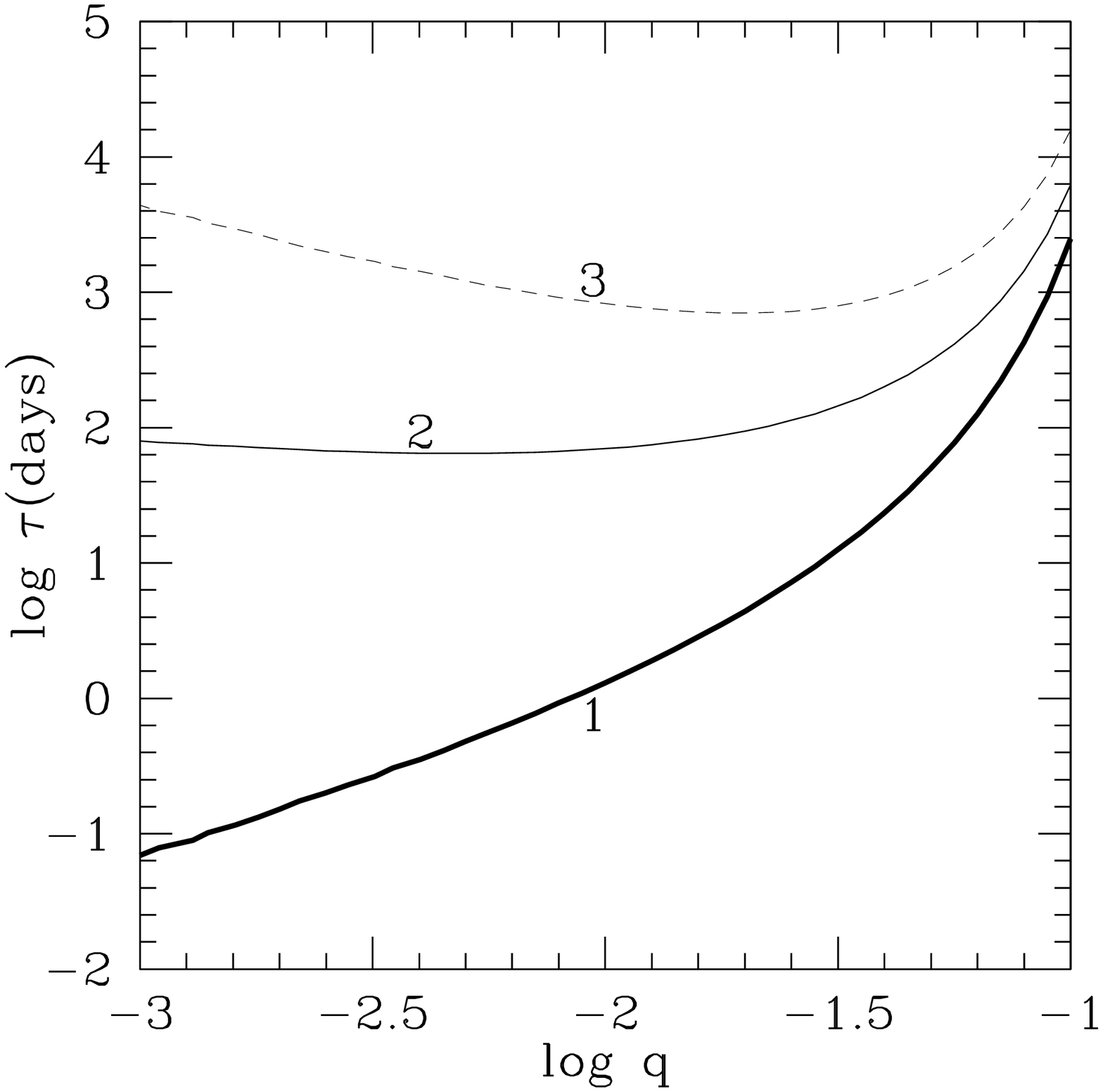}}
\caption{The dissipated luminosity (left) and the time scale (right)
of the violent merger
as a function of the mass ratio, $q$, for three cases of the primary star,
as described in the text.
Full thick curves: case 1 ($1\,M_\odot$ main sequence).
Full thin curves: case 2 ($8\,M_\odot$ main sequence).
Dashed curves: case 3 ($3\,M_\odot$ red giant).
}
\label{lum_tau}
\end{figure}

Results of substituting the above parameters to Eqs. (\ref{lum_d}) and
(\ref{tau1}) are shown in Fig.~\ref{lum_tau}. In all the cases $f = 0.9$ has
been assumed in Eq.~(\ref{tau1}). The above described cases 1, 2, and 3 are
presented with full thick, full thin, and dashed curves, respectively.

As can be seen from Fig. \ref{lum_tau} the mergers with the main sequence
stars are more energetic, in terms of the dissipated luminosity, as those
with a red giant. This is due to lower densities in the giant envelope and
larger radii, hence lower velocity differences, at which the energy
dissipation takes place.

In the case of a solar type star the merger with a low mass object gives
rise in luminosity up to $\sim 10^6\,L_\odot$ but on a short time scale. In
the case of a Jupiter-like planet this would be a burst as short as an hour.
The burst would however be followed by a decline on a much longer time scale.

The case of the $8\,M_\odot$ main-sequence star is relevant to the outburst
of V838~Mon (TS06). For a mass ratio $q \simeq 0.03$
Fig.~\ref{lum_tau} predicts a luminosity of a few times $10^5\,L_\odot$
on a time scale of a 100~days, which is in a reasonable agreement with the
observations (see, e.g. Tylenda 2005). Note however that in order do get a
better agreement with the observations TS06 propose that
the merger in V838~Mon occurred from a highly eccentric orbit.

In the case of a red giant star the merger event is relatively mild. It can
rise the stellar luminosity by a factor of a few only but on a time scale as
long as several years.

\section{DISCUSSION}

In Section 1 we distinguish between three basic
types of merger events:
(1) A violent event where the secondary is disrupted and supplies most
of the ejected and inflated envelope mass. This occurs when the secondary
star is loosely bound, e.g., a pre-main-sequence star, and there is a grazing
(rather than a head-on) collision from an eccentric orbit.
(2) A violent merger where the secondary survives to a deeper depth in the
primary envelope, and most of the inflated envelope originates in the
very outer layers of the primary stellar envelope.
Such a merger is likely to occur in a binary system when the secondary cannot
maintain the primary envelope in synchronization (corotation) with the orbital motion.
(3) A non-violent merger, occurring when the secondary maintains
synchronization until merger occurs.

The basic results of the paper is that the type-2 merger defined above
can cause a very bright event when the primary is a main sequence star.
The peak luminosity can reach $L_{\rm peak} \ga 10^5 L_\odot$, and last up to a few months.
Our claim might seem somewhat contrary to intuition, as one might expect the
merger event to become more violent as the secondary mass increases.
However, as the secondary stellar mass increases, the relative velocity between
the secondary star and the primary stellar envelope at the moment of merger decreases,
implying weaker shock waves in the very outer layers of the envelope.

Interestingly, we find that the type-2 merger can lead to a merger
event similar to that observed in V838 Mon, although type-1 event seems
to fit observations a little better (TS06; Soker \& Tylenda 2007).
The hugely inflated envelope in the type-1 merger event
discussed by TS06 for V838 Mon, must have both a grazing collision
and a pre-main-sequence secondary star.
These ingredients are not required in the type-2 merger scenario.
However, the type-2 merger scenario cannot lead to a massive inflated envelope.
The total mass in the expelled shell and in the inflated envelope of
V838 Mon is $M \simeq 0.1-0.3 M_\odot$ (Tylenda 2005; Soker \& Tylenda 2007).
This mass favors the type-1 merger for V838 Mon. However, if this
mass will turn out to be $M \la 0.05 M_\odot$ the type-2 merger model
might work as well.

One of the important observational aspects of the merger events is the
decline phase. Mass loss from a merger is always much smaller than the
mass disturbed in the event. Most of this disturbed matter will form a more
or less inflated envelope of the primary. When the merger processes
dissipating energy are over
the only source of luminosity in the envelope
is the gravitational energy released in its contraction. However, the
thermal time scale of the envelope is comparable to or lower than its dynamical
time scale, especially if the envelope outer radius is much larger than the
thermal equilibrium radius of the primary. Therefore the envelope contraction phase
will be proceeded by a rapid cooling of the envelope outer layers. This
cooling can go down to the Hayashi limit. Therefore the merger remnant is
expected to decline as a very cool star. This is the main observational
aspect allowing to distinguish the merger events from thermonuclear runaway
events, e.g. nova-type outbursts. In the latter case, as it is well known
from theoretical models and observations, the object has to evolve to
very high effective temperatures ($\ga 10^5$\,K) before the final decline.

{{{ The exact evolution of the merger remnant as it re-establishes equilibrium
requires numerical calculations. Podsiadlowski (2003) has
considered an instantaneous removal of mass from a subgiant, together with
an instantaneous addition of energy to its remaining envelope.
When the heating is sufficiently high, in a short time the
star reaches high
luminosity, and then it contracts more or less along the Hayashi line.
However, when a mass is removed and the heating
(energy addition) is not sufficient, the star becomes very underluminous
(Podsiadlowski 2003).
The less-heated star starts its life to the left of the
Hayashi line on the HR diagram, and then contracts and fades, but as
a less luminous and smaller star than a star of the same parameters on the
Hayashi line.
The reason is that after mass is instantaneously removed from the outer
layers of the
envelope deep layers have to expand outward. This requires a lot of energy.
In the merger process mass is added and lots of orbital energy is
released.
Thus there is no need for the inner envelope's layers to expand.
On the contrary, after the short outburst which makes the star luminous and
inflates an envelope, layers in the extended envelope have to contract.
We therefore expect the merger remnant to decline along the Hayashi line.
A behavior like this was observed during the eruption and subsequent
decline of V838 Mon (Tylenda 2005). }}}

Soker \& Tylenda (2007) suggest to call the violent merger events
{\it mergeburst.} As it is clear from our present study these events can be
easily observable not only in our Galaxy but also beyond it.
What is the galactic {\it mergeburst} rate?
We can crudely estimate it as follows.
We use the estimated formation rate of blue stragglers as done by
Ciardullo et al. (2005) to explain bright planetary nebulae in elliptical galaxies.
Ciardullo et al. (2005) argue that blue stragglers can account for the formation
of bright planetary nebulae in elliptical galaxies with an old stellar population,
and estimate the blue stragglers formation rate per solar luminosity in galaxies.
Their estimates yields a galactic blue stragglers formation rate of
$\sim 0.1 \yr^{-1}$.
The blue stragglers in their scenario requires that the secondary mass be
close to that of the primary, $0.7 M_1 \la M_2 <M_1$, whereas in our scenario
$0.003 M_1 \la M_2 \la 0.15 M_1$.
Because the allowed range in $M_2$ is smaller in our proposed model, and the mass ratio
is much smaller, we expect the type-2 {\it mergeburst} rate to be less frequent,
say $0.01-0.05 \yr^{-1}$.
On the other hand, the merger rate of young systems due to perturbation
from the environment, as proposed for V838 Mon (TS06), can add to
the number of bright transient {\it mergeburst} events.
Over all, we estimate that bright transient {\it mergeburst} events,
resulting from stellar merger, will occur in the galaxy once every $\sim 10-50$~yr.

\acknowledgments
We thank the referee, Philipp Podsiadlowski, for useful comments.
This research was supported in part by the Asher
Fund for Space Research at the Technion, as well as, from the Polish State
Committee for Scientific Research grant no. 2~P03D~002~25.

\end{document}